\documentstyle[12pt]{article}
\def\ie{{\em i.e.}}
\def\eg{{\em e.g.}}

\def\beq{\begin{equation}}
\def\eeq{\end{equation}}

\catcode`\@=11 
\def\coeff#1#2{{\textstyle{#1\over #2}}}

\def\lsim{\mathrel{\mathpalette\@versim<}}
\def\gsim{\mathrel{\mathpalette\@versim>}}
\def\@versim#1#2{\vcenter{\offinterlineskip
    \ialign{$\m@th#1\hfil##\hfil$\crcr#2\crcr\sim\crcr } }}
\def\etal{{\em et. al.}}
\def\JL{J. L. Lopez}
\def\DVN{D. V. Nanopoulos}
\def\AZ{A. Zichichi}

\def\t1{{\tilde 1}}

\def\GeV{\,{\rm GeV}}

\def\to{\rightarrow}

\def\NPB#1#2#3{Nucl. Phys. B {\bf#1} (19#2) #3}
\def\PLB#1#2#3{Phys. Lett. B {\bf#1} (19#2) #3}

\def\PRD#1#2#3{Phys. Rev. D {\bf#1} (19#2) #3}
\def\PRL#1#2#3{Phys. Rev. Lett. {\bf#1} (19#2) #3}

\def\TAMU#1{Texas A \& M University preprint CTP-TAMU-#1}

\begin{document}
\begin{flushright}
\baselineskip=12pt
{CERN-TH.7520/94}\\
{CTP-TAMU-63/94}\\
{ACT-21/94}\\
{hep-ph/9412237}\\
\end{flushright}

\begin{center}
\vglue 0.5cm
{\Huge\bf Direct detection of dark matter in SU(5)$\times$U(1) supergravity\\}
\vglue 1cm
{XU~WANG$^{1,2}$, JORGE L. LOPEZ$^{1,2}$, and D. V. NANOPOULOS$^{1,2,3}$\\}
\vglue 0.4cm
{\em $^{1}$Center for Theoretical Physics, Department of Physics, Texas A\&M
University\\}
{\em College Station, TX 77843--4242, USA\\}
{\em $^{2}$Astroparticle Physics Group, Houston Advanced Research Center
(HARC)\\}
{\em The Mitchell Campus, The Woodlands, TX 77381, USA\\}
{\em $^{3}$CERN Theory Division, 1211 Geneva 23, Switzerland\\}
\baselineskip=12pt
\end{center}

\vglue 0.5cm
\begin{abstract}
We compute the scattering rates for the lightest neutralino $\chi^0_1$ in
the forthcoming germanium $(^{73}{\rm Ge}+^{76}{\rm Ge})$ detector and a
proposed lead detector $(^{207}{\rm Pb})$,  within the framework of
$SU(5)\times U(1)$ supergravity. We find that in only a small portion
($\lsim10\%$) of the parameter spaces of this class of models, are the rates
in the germanium detector above the expected initial experimental sensitivity
of 0.1 events/kg/day. However, a much larger portion ($\lsim40\%$) of the
parameter spaces could be probed with an improved background rejection
capability (0.01 events/kg/day) and/or a more sensitive detector $(^{207}{\rm
Pb})$.
\end{abstract}

\vspace{1cm}
\begin{flushleft}
\baselineskip=12pt
{CERN-TH.7520/94}\\
{CTP-TAMU-63/94}\\
{ACT-21/94}\\
November 1994
\end{flushleft}
\newpage

\setcounter{page}{1}
\pagestyle{plain}
\baselineskip=14pt

\section{Introduction}
The search for supersymmetry is being carried out in many ways. In high energy
collider experiments, such as the Tevatron and LEP, various measurements have
yielded lower bounds on the sparticle masses, although no signal has yet been
observed. It is possible that we might need to wait for the next generation of
colliders to further probe the spectrum of supersymmetric particles, although
we may be able to get some interesting results through other types of
experiments, such as the direct detection of dark matter particles. It is
well-known that in supersymmetric theories where R--parity is conserved, the
lightest supersymmetric particle (LSP) is stable, and it is a candidate for
cold dark matter if neutral and colorless \cite{EHNOS}, as the lightest
neutralino is $(\chi\equiv\chi^0_1)$. According to astronomical observations,
most notably from the refurbished Hubble Space Telescope, we know that most of
the matter in the universe is invisible. If we define
$\Omega_{\chi}=\rho_{\chi}/\rho_{c}$, where $\rho_{c}$ is the critical density
to ``close" the universe and $\rho_{\chi}$ the neutralino dark matter density,
it is known that $\Omega_{\chi} h^{2}\sim 0.5$ where $h = H/(100 {\rm
km\,s^{-1}\,Mpc^{-1}})$ and $H$ is the Hubble parameter. Since recent Hubble
Telescope observations indicate that $h\simeq 0.8$ \cite{Freedman}, we see that
$\Omega_{\chi}\sim1$ is expected.

With this large amount of dark matter in the universe, it is expected that we
should be able to detect the recoiling effect of the neutralino, as a dark
matter particle, scattering off the target nuclei of certain types of
detectors. Indeed, some experimental effort have been put into reality in this
regard \cite{experiment}. We consider two detectors
$^{73}{\rm Ge} (500 g)+^{76}{\rm Ge} (500 g)$ and $^{207}{\rm Pb}$. The Ge
detector is currently being set up at Stanford University, whereas the second
one is still being discussed. There are two points regarding the detector
dependence of the scattering processes which are worth mentioning: (i) There
are two processes through which the $\chi^0_1$ interacts with the partons in
the protons and neutrons inside the target nucleus. The first one is the
spin-dependent
(S.D.) incoherent scattering, which is non--zero only when the target nucleus
has non-zero spin. The second one is the spin-independent coherent scattering
which increases with the mass of the target nucleus ($\propto m_N$). These two
processes do not interfere with each other, thus the total rate is just the
direct sum of these two rates. For the two detectors considered here the
spin-dependent contribution (S.D.) is always negligible compared with the
spin-independent (S.I.) contribution. (ii) Because of these features, a
detector with the highest possible atomic number is prefered: $^{207}{\rm Pb}$
is the superconducting material with highest atomic number. Concerning the size
of
the detector, one would naively think that since one wants to have the largest
event rate, one should build the largest possible detector. However, this is
not the case with the cryogenic experimental set up, whose operation is
optimized for a 1kg detector mass.

\section{The models and their parameter space}
\label{models}
In what follows we investigate the parameter space of no-scale $SU(5)\times
U(1)$ supergravity. For a recent review on the construction of these models
see Ref.~\cite{faessler}. The effect of several other direct and indirect
experimental constraints have been discussed in Refs.~\cite{Easpects,One}.
These models have the features of universal soft-supersymmetry-breaking at the
unification scale and radiative breaking of the electroweak symmetry, which is
enforced by minimizing the one-loop effective potential. Generally speaking,
there are four free parameters that describe these models, once the top-quark
mass is fixed. They can be chosen to be: $m_{\chi^{\pm}_1}, \xi_{0}, \xi_{A},
\tan\beta$, where $m_{\chi^\pm_1}$ is the lightest chargino mass, $\tan\beta$
is the ratio of Higgs vacuum expectation values, and $\xi_{0}\equiv
m_0/m_{1/2}$ and $\xi_{A}\equiv A/m_{1/2}$ are ratios of the usual universal
soft-supersymmetry-breaking parameters.
In string-inspired supersymmetry breaking scenarios two parameters are
eliminated: in the {\em moduli} scenario $m_{0} = A = 0$, whereas in the {\em
dilaton} scenario $m_{0} = \coeff{1}{\sqrt{3}}m_{1\over 2}, A = -m_{1\over 2}$.
In either case there are only two free parameters left $(m_{\chi^{\pm}_{1}},
\tan\beta)$. We also consider the ``strict'' version of these scenarios where
the bilinear scalar coupling $B$ is also specified at the unification scale: in
the (strict) moduli scenario $B(M_{U}) = 0$, and in the (strict) dilaton
$B(M_{U}) = 2m_{0}$. In this case the value of $\tan\beta$ can also
be determined, and we are left with one-parameter models. Compared with generic
low-energy supersymmetric models like the Minimal Supersymmetric Standard Model
(MSSM), these few-parameter models are better motivated and much more
predictive.

The main inputs from these models to the dark matter detection calculation are:
the lightest CP-even Higgs mass $m_{h}$, and the lightest neutralino mass
$m_{\chi^0_1}$. The allowed ranges for these masses after all model constraints
have been imposed are: $m_{h} \simeq (65-115)\GeV$ and $m_{\chi^{0}_1} \simeq
(23-145)\GeV$. An important mass relation that one should keep in mind
throughout our discussion is $m_{\chi^0_1}\approx{1\over2}m_{\chi^\pm_1}$.

\section{The scattering rate}
\label{formula}
\begin{figure}[p]
\vspace{6in}
\includegraphics{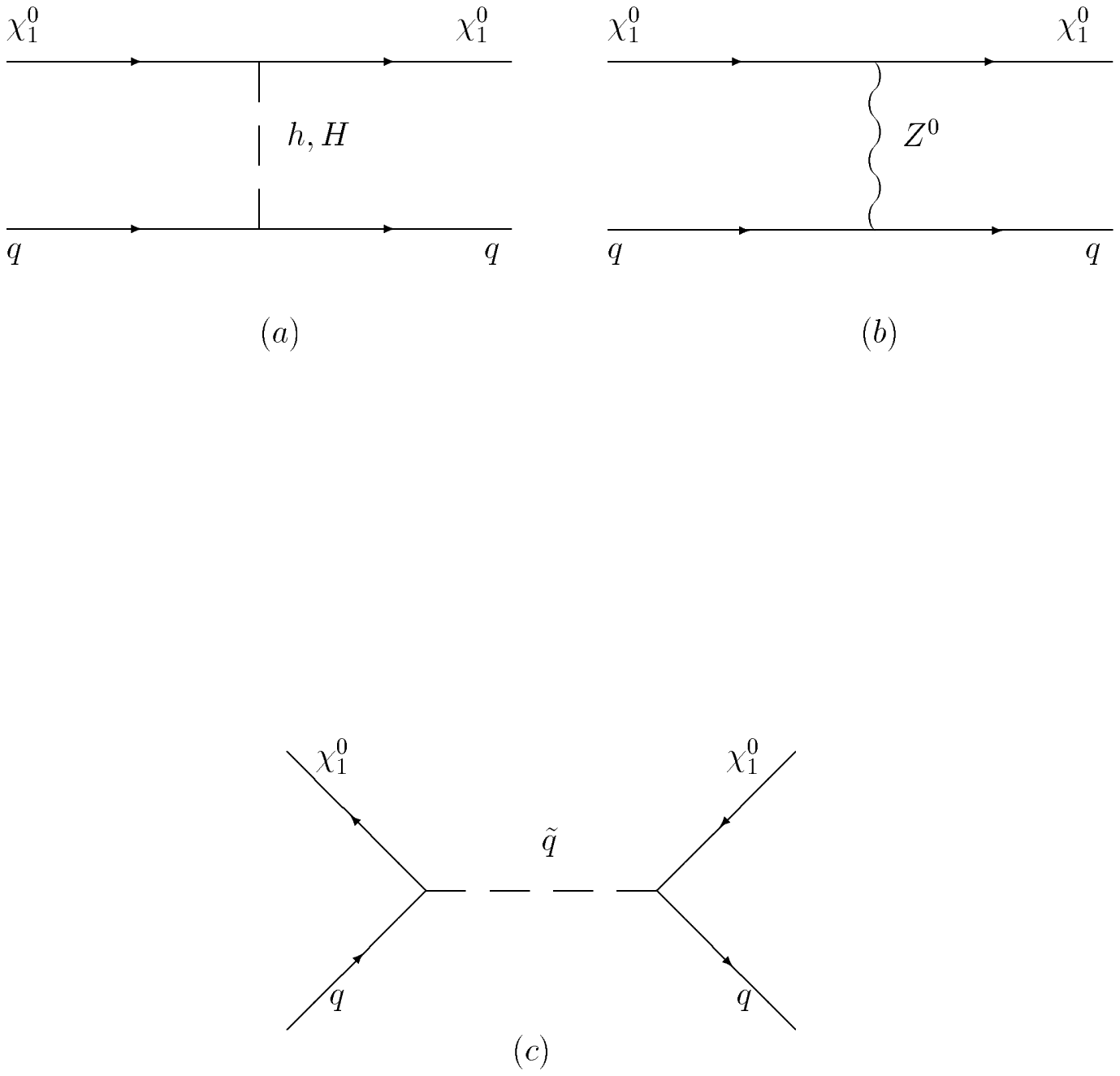}
\caption{The Feynman diagrams for the elastic $\chi^0_1$-quark scattering
process, which occurs via $t$-channel $h,H$- and $Z$-boson exchange $(a, b)$
and via $s$-channel $\tilde q$-exchange $(c)$.}
\label{lspd_feynman}
\end{figure}

The Feynman diagrams for the elastic scattering between the $\chi_1^0$ and
the partons inside the nucleons in the target nucleus ($\chi_1^0 + q \to
\chi_1^0 + q$) are shown in Fig.~\ref{lspd_feynman}.
In order to calculate the scattering rate
we need to write down the effective Lagrangian for this process. If
we assume that most of the dark matter halo surrounding our galaxy is made
of elementary particle dark matter, then the typical dark matter velocity
is $v \approx 300{\rm km/s} = 10^{-3}c$, where $c$ is the speed of light.
At these velocities there is an energy transfer $\Delta E < m_{\chi_1^0}v^2  =
10 {\rm keV} (m_{\chi_1^0}/10\GeV)$ which is much smaller than the mass scale
involved in the process ($m_{\chi_1^0}$). Therefore, we can use the effective
Lagrangian approach to describe the low-energy $\chi^0_1$-quark interaction,
\begin{equation}
L_{eff} = \bar{\chi_1^0}\gamma_{\mu}\gamma_{5}\chi_1^0\cdot
\bar q\gamma^{\mu}(A_qP_L+B_qP_R)q
+ \bar{\chi_1^0}{\chi_1^0}\cdot {C_{q}}m_{q}q
\end{equation}
where $P_{R(L)} = \coeff{1}{2}(1\pm\gamma_{5})$ and $A_q, B_q$ ($q=u,d,s$)
are the coupling coefficients for the $Z$-exchange in the $t$-channel and the
$\tilde q$-exchange in the $s$-channel; only light quarks contribute in this
case. Also, $C_q\ (q=u,d,s,c,b,t)$ are the coupling coefficients for the
$h,H$-exchange $t$-channel and the $\tilde q$-exchange $s$-channel. Note that
heavy quarks do not decouple in this case. Among the various contributions to
the scattering amplitude, the squark contribution is negligible since
$m_{\tilde q} > 200\GeV$ in these models. The dominant contribution is from
spin-independent coherent scattering due to (CP-even) Higgs boson exchange,
mostly the lightest one ($h$) although the heavy Higgs ($H$) can have a
significant sub-leading effect.

The total scattering rate is given by \cite{formula}
\begin{equation}
R = (R_{\rm S.I.}+R_{\rm S.D.}){{4m_{\chi_1^0}m_{N}}\over {(m_{\chi_1^0}+
m_{N})^{2}}}\,{{\rho_{\chi}}\over {0.3\GeV {\rm cm}^{-3}}}\,{{|V_{E}|}\over
{320{\rm kms}^{-1}}}\,{{{\rm events}}\over {{\rm kg}\cdot {\rm day}}}\ ,
\label{R}
\end{equation}
where $m_{N}$ is the target nucleus mass, $\rho_{\chi}$ is the
neutralino dark matter relic density, $|V_{E}|$ is the average velocity of
the neutralinos that hit the detector (we take $|V_E|=320\,{\rm km\,s^{-1}}$),
and
\begin{equation}
R_{\rm S.I.}=840m_{N}^{2}M_{Z}^4\zeta\left[\hat f\,{m_uC_u+
m_dC_d\over m_u+m_d}+fC_s+\coeff{2}{27}(1-f-\hat f)(C_c+C_b+C_t)\right]^{2}\ .
\end{equation}
In the expression for $R_{\rm S.I.}$,
$\zeta={0.573\over b}[1-{e^{-b/(1+b)}\over\sqrt{1+b}}\,{{\rm erf}
({1\over\sqrt{1+b}})\over{\rm erf}(1)}]$ is a form factor with $b={m^2_\chi
m^2_N\over(m_\chi+m_N)^2}{8\over9}\sigma^2r^2_{\rm charge}$,
$\sigma={V_E\over1.2}$ the velocity dispersion, and the charge radius $r_{\rm
charge}=(0.30+0.89A^{1/3})\,{\rm fm}$. Also, $\hat f\approx0.05$ and
$f\approx0.2$ are coefficients of proportionality in various hadronic matrix
elements. Finally, the $C_q$ parameters for the dominant Higgs exchange
diagrams involve couplings of the Higgs to neutralinos and to quarks:
\begin{equation}
C_q=-{g^2_2\over4M^2_W}\left\{\begin{array}{c}{-\cos\alpha\over\sin\beta}\\
{\sin\alpha\over\cos\beta}\end{array}\right\}{F_h\over m^2_h}
+{g^2_2\over4M^2_W}\left\{\begin{array}{c}{\sin\alpha\over\sin\beta}\\
{\cos\alpha\over\cos\beta}\end{array}\right\}{F_H\over m^2_H}\ ;\quad
\left\{\begin{array}{c}q=u,c,t\\ q=d,s,b\end{array}\right.\ ,
\label{Cq}
\end{equation}
where $\alpha$ is the Higgs mixing angle, $F_h=(N_{11}-N_{12}\tan\theta_W)
(N_{14}\cos\alpha+N_{13}\sin\alpha)$, $F_H=(N_{11}-N_{12}\tan\theta_W)
(N_{14}\sin\alpha-N_{13}\cos\alpha)$, and the LSP is the admixture
$\chi^0_1=N_{11}\widetilde W^3+N_{12}\widetilde B+N_{13}\widetilde H^0_1+N_{14}
\widetilde H^0_2$.
It can be seen that $R_{\rm S.I.}\propto m^2_N$, thus the reason for proposing
the $^{207}{\rm Pb}$ detector. Also, explicit calculations show that for both
$^{73}{\rm Ge}+^{76}{\rm Ge}$ and $^{207}{\rm Pb}$ detectors, the S.I. term is
always much larger
than the S.D. term (which we do not exhibit explicitly here).

\begin{table}[t]
\begin{center}
\caption{The fraction of parameter space in $SU(5)\times U(1)$ supergravity
which can be explored via direct detection of dark matter in Ge and Pb
detectors with a sensitivity of 0.1 or 0.01 events/kg/day.}
\label{table}
\smallskip
\begin{tabular}{|c||c|c||c|c||c||c||c|c|}\hline
&\multicolumn{2}{c||}{Moduli}&\multicolumn{2}{c||}{Dilaton}&\multicolumn{1}{c||}{Strict Moduli}&\multicolumn{1}{c||}{Strict Dilaton}\\ \hline
&$\mu>0$&$\mu<0$&$\mu>0$&$\mu<0$&$\mu<0$&$\mu<0$\\ \hline
Ge(0.1)&2.1\%&7.1\%&10\%&11\%&10\%&5.2\%\\
Ge(0.01)&14\%&30\%&32\%&39\%&41\%&29\%\\
Pb(0.1)&3.6\%&11\%&18\%&21\%&17\%&16\%\\
Pb(0.01)&27\%&49\%&43\%&53\%&59\%&45\%\\ \hline
\end{tabular}
\end{center}
\hrule
\end{table}

\section{Results and discussion}
\begin{figure}[p]
\vspace{6in}
\includegraphics{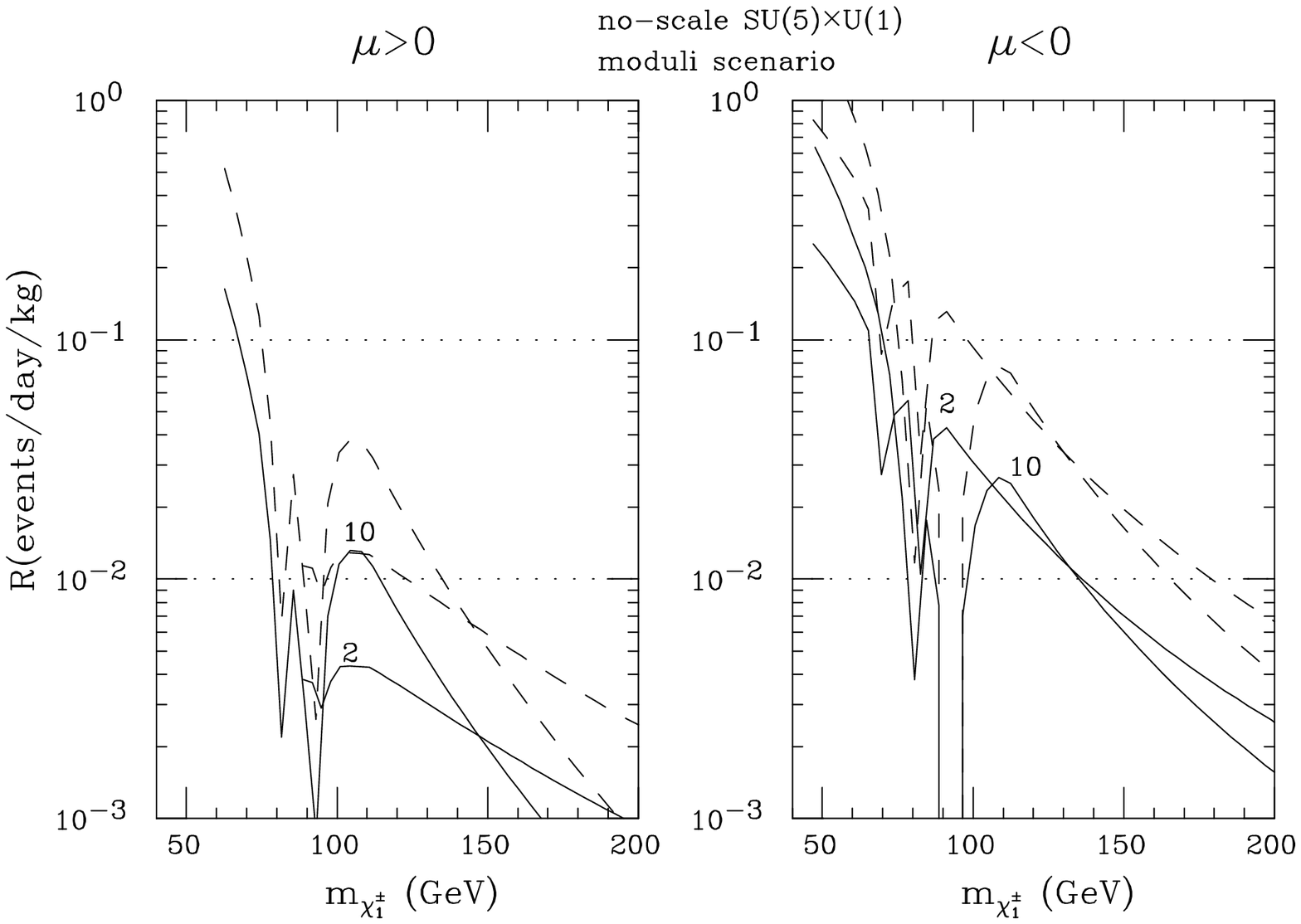}
\caption{The calculated scattering rates for the Ge (solid lines) and
Pb (dashed lines) dark matter detectors in no-scale
$SU(5)\times U(1)$ supergravity -- moduli scenario for $\tan\beta=2,10$. The
horizontal dotted lines represent the detector sensitivities of 0.1 and 0.01
events/kg/day. The dips in the rates correspond to a suppressed neutralino
relic density when the $Z$ and $h$ poles in the neutralino pair annihilation
are encountered.}
\label{lspd_nsc}
\end{figure}

\begin{figure}[p]
\vspace{6in}
\includegraphics{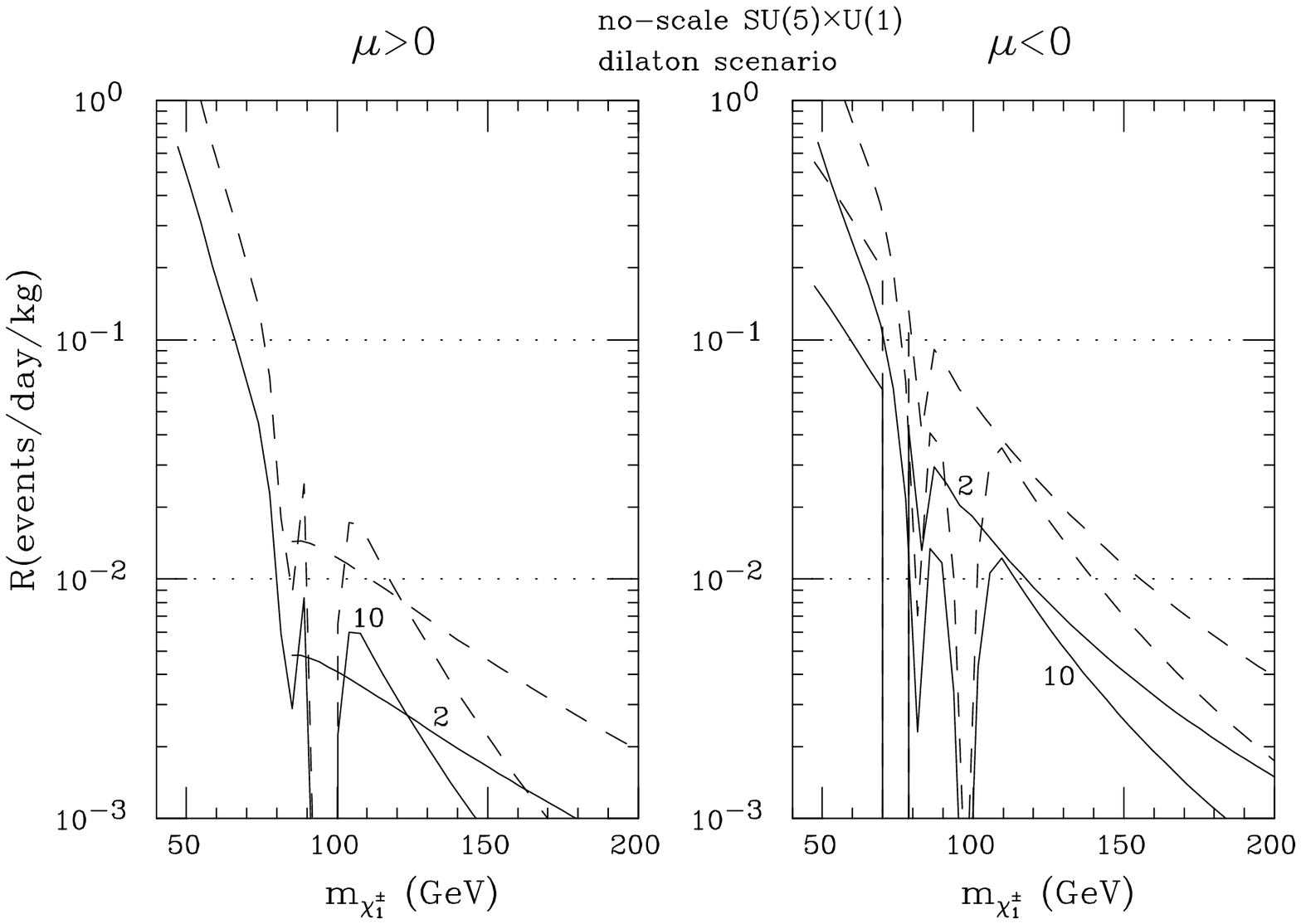}
\caption{The calculated scattering rates for the Ge (solid lines) and
Pb (dashed lines) dark matter detectors in no-scale
$SU(5)\times U(1)$ supergravity -- dilaton scenario for $\tan\beta=2,10$. The
horizontal dotted lines represent the detector sensitivities of 0.1 and 0.01
events/kg/day. The dips in the rates correspond to a suppressed neutralino
relic density when the $Z$ and $h$ poles in the neutralino pair annihilation
are encountered.}
\label{lspd_kl}
\end{figure}

\begin{figure}[p]
\vspace{6in}
\includegraphics{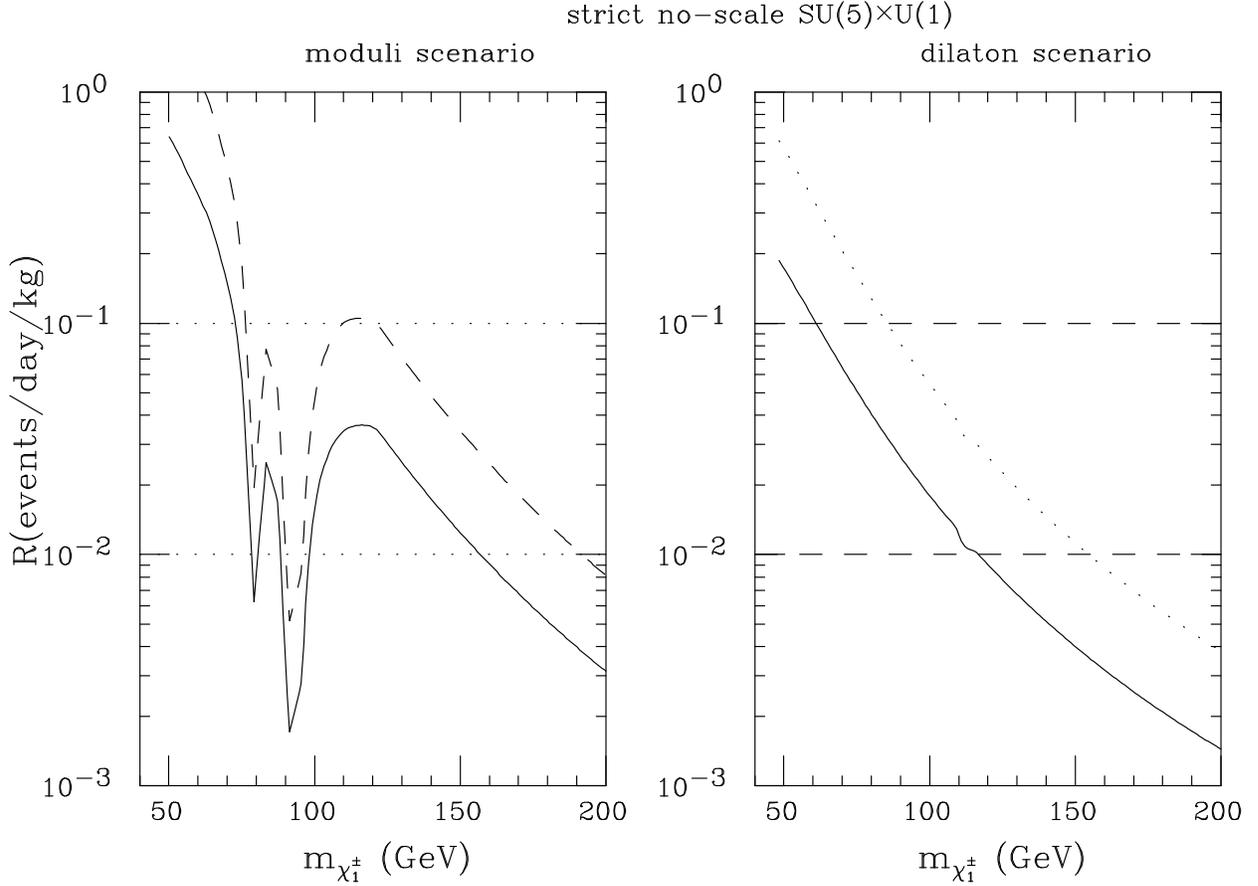}
\caption{The calculated scattering rates for the Ge (solid line) and Pb
(dashed line) dark matter detectors in strict no-scale
$SU(5)\times U(1)$ supergravity -- moduli and dilaton scenarios.
The dotted lines represent the detector sensitivities of 0.1 and 0.01
events/kg/day. The dips in the rates correspond to a suppressed neutralino
relic density when the $Z$ and $h$ poles in the neutralino pair annihilation
are encountered.}
\label{lspd_strict}
\end{figure}

We have calculated the scattering rates for the neutralino dark matter
particles in the two detectors mentioned above throughout the parameter space
of the models. The results are shown in the Figs.~\ref{lspd_nsc},
\ref{lspd_kl}, \ref{lspd_strict} for the different scenarios. The following
general features are noticed in all cases:
\begin{itemize}
\item The rates drop quickly with growing chargino masses. As mentioned above,
the lightest neutralino is a linear combination of gaugino and higgsino gauge
eigenstates. The intermediate states for the dominant spin-independent
scattering process are the Higgs bosons, which couple to a product of the
gaugino and higgsino components of the neutralino (see Eq.~(\ref{Cq})). On the
other hand, when the chargino mass grows the neutralino approaches the limit of
a pure bino, thus greatly reducing its coupling to the Higgs bosons, and the
scattering rate drops sharply.
\item There are two noticeable dips in the rates for
$m_{\chi^\pm}\sim70-90\GeV$. For
$m_{\chi^0_1}\approx{1\over2}m_Z,{1\over2}m_h$, in the calculation of the relic
density of the neutralinos (which we perform following the methods of
Ref.~\cite{dmcalc}), the $Z$ or $h$-pole is encountered in the neutralino
pair-annihilation, and the resulting neutralino relic density is highly
suppressed. This implies that little neutralino dark matter is expected in the
halo. To account for this possibility, in Eq.~(\ref{R}) we take
\begin{equation}
\rho_{\chi} = {\rm min}\,[1,(\Omega_\chi h^2)/{0.05}]\, 0.3
{\GeV}{\rm cm}^{-3}\ ,
\end{equation}
since one expects that $\Omega_{\rm halo}\gsim0.1$ \cite{KT}. In the figures,
the ordering of the poles with increasing $m_{\chi^\pm_1}$ is $h,Z$ for
$\tan\beta=2$, and $Z,h$ for $\tan\beta=10$ (and in Fig.~\ref{lspd_strict}).
\item The rates have a non-trivial $\tan\beta$ dependence (only
$\tan\beta=2,10$ are shown in the figures). The $C_q$ coefficients in
Eq.~(\ref{Cq}) depend on $\tan\beta$, and so does $m_h$, which increases with
$\tan\beta$.
\end{itemize}

For the Ge detector, which is the one being set up right now and should start
producing data in the near future, with the realistic experimental sensitivity
of 0.1 event/kg/day, only a small portion ($\sim2-10\%$) of the parameter space
would be probed in all scenarios (see Table~\ref{table}). These fractions have
been determined by computing the rates for {\em all} points in parameter space;
only selected values of $\tan\beta$ are shown in the figures. There are two
ways to improve the reach into parameter space. First, one could improve the
sensitivity by reducing the background. If the sensitivity of the Ge detector
could be enhanced to 0.01 event/kg/day, a larger fraction of the parameter
space could be probed ($\sim15-40\%$), as seen in the figures and quantified in
Table~\ref{table}.

There has also been some discussions about the possibility of building a
$^{207}{\rm Pb}$ detector, which is an improvement over the Ge detector because
the scattering rate scales about linearly ($R\propto m_N/(1+m_\chi/m_N)^2$)
with the atomic number of the target nucleus, \ie, an improved sensitivity by
increasing the signal in the detector. With the 0.1 event/kg/day sensitivity
the Pb detector should probe a larger ($\sim\times2$) fraction of the parameter
space, compared to the Ge detector with the same sensitivity (see
Table~\ref{table}). Of course the best experimental scenario would be a
$^{207}{\rm Pb}$ detector with the better sensitivity (0.01 events/kg/day). In
such best case scenario, for $\tan\beta=2$ and $\mu<0$, chargino masses as high
as $180\,(160)\GeV$ could be probed in the moduli (dilaton) scenario. The reach
is somewhat lower for larger values of $\tan\beta$. Note also that in strict
no-scale $SU(5)\times U(1)$ -- dilaton scenario, an experimental upper bound on
$R$ would immediately provide a lower bound on $m_{\chi^\pm_1}$.

It has been recently pointed out \cite{BDN} that experimental limits on
$B(b\to s\gamma)$ may disfavor some regions of parameter space where $R$ is
particularly enhanced. The computations of $B(b\to s\gamma)$ in supergravity
models in Ref.~\cite{LargeTanB}, in particular for no-scale $SU(5)\times U(1)$
supergravity, indicate that $b\to s\gamma$ is an important constraint only for
$\mu>0$, and even in that case only for $\tan\beta\gsim4$. Therefore, even
assuming that the $b\to s\gamma$ constraints are rigorous, there should still
be a large portion of the parameter space accessible to direct searches of
dark matter.

We conclude with the following comment. At LEPII, chargino masses as high as
100 GeV should be readily detectable in all the scenarios which we have
considered here \cite{Easpects,One}, and yet, direct detection of the LSP with
$m_{\chi^0_1}\approx{1\over2}m_{\chi^\pm_1}\lsim50\GeV$ may take a while to
occur. Indirect detection of dark matter via upwardly moving muon fluxes in
neutrino telescopes may taken even longer to happen, as explicit calculations
show in the models we considere here \cite{NT} and more generally in
model-independent analyses \cite{KGJS}.

\section*{Acknowledgments}
This work has been supported in part by DOE grant DE-FG05-91-ER-40633. The
work of X. W. has been supported by the World Laboratory. We would like to
thank D. Caldwell for information on the status of the experiments.

\newpage

\clearpage

\clearpage

\clearpage

\clearpage

\clearpage

\end{document}